\documentclass[10pt,a4paper]{article}
\usepackage{amsfonts,latexsym,amssymb,graphicx}

\makeatletter
\def\bs{\ensuremath{\backslash}}
\def\XOR{\;{\operator@font XOR}\;}
\def\atn{\mathop{\operator@font atn}\nolimits}
\def\Reals{\mathbb{R}}

\def\czp#1{\hbox to 0pt{\hss #1\hss}}  
\def\majordots{\mathinner{\mkern1mu \raise8pt\vbox{\kern7pt\hbox{.}} \mkern-1.8mu \raise 4pt\hbox{.} \mkern-1.8mu \raise0pt\hbox{.} \mkern1mu}}
\def\minordots{\mathinner{\mkern1mu \raise0pt\hbox{.} \mkern-1.8mu \raise 4pt\hbox{.} \mkern-1.8mu \raise8pt\vbox{\kern7pt\hbox{.}} \mkern1mu}}

\def\suchthat{\,:\,}
\def\from{\mathchar"3220\relax}

\def\of{{{}_\circ}}\def\of{\relax}
\def\ox{\otimes}

\def\ket#1{{\left|#1\right\rangle}}
\def\proj#1{{\left|#1\right\rangle\left\langle#1\right|}}
\def\braopket#1#2#3{\left\langle#1\vphantom{#2#3}\right|#2\left|\vphantom{#1#2}#3\right\rangle}

\long\def\ignore#1{}

\ignore{
\newif\if@pars
\DeclareRobustCommand\cite{%
  \@parsfalse\@ifnextchar [{\@tempswatrue\@citex}{\@tempswafalse\@citex[]}}
\DeclareRobustCommand\parcite{%
  \@parstrue\@ifnextchar [{\@tempswatrue\@citex}{\@tempswafalse\@citex[]}}
\def\@citex[#1]#2{%
  \let\@citea\@empty%
  \def\shift@smaybe{\gdef\@smaybe{}\def\shift@smaybe{\gdef\@smaybe{s}}}%
  \@cite{\@for\@citeb:=#2\do
    {\@citea\def\@citea{,\penalty\@m\ }\shift@smaybe%
     \edef\@citeb{\expandafter\@firstofone\@citeb\@empty}%
     \if@filesw\immediate\write\@auxout{\string\citation{\@citeb}}\fi
     \@ifundefined{b@\@citeb}{\mbox{\reset@font\bfseries ?}%
       \G@refundefinedtrue
       \@latex@warning
         {Citation `\@citeb' on page \thepage \space undefined}}%
       {\hbox{\csname b@\@citeb\endcsname}}}}{#1}}
\def\@cite#1#2{{\setbox0\hbox{#1}\if@pars(\fi{Ref\@smaybe.~#1\if@tempswa , #2\fi}\if@pars)\fi}}
}  
\let\parcite\cite

\def\@maketitle{{\def\baselinestretch{1.3}%
 \newpage\null\vskip2em\begin{center}%
  \let\footnote\thanks%
  {\LARGE\@title\par}\vskip1.5em{\large\lineskip.5em\begin{tabular}[t]{c}\@author\end{tabular}\par}%
  \vskip1em{\large\@date}%
 \end{center}\par\vskip0em%
}}

\newenvironment{pseudocode}[1]{\vskip12pt plus3pt minus3pt\noindent{\normalsize\bf #1}\def\	{\hskip30pt}\catcode`\
=\active}{\vskip12pt plus3pt minus3pt}
\makeatother

\begin{document}

\title{Quantum basin hopping with gradient-based local optimisation}
\author{David Bulger\thanks{Lecturer, Department of Statistics, Macquarie University, NSW 2109, Australia,
 {\tt dbulger@efs.mq.edu.au}.}}

\maketitle

\thispagestyle{empty}

\begin{abstract}
The quantum basin hopping algorithm for continuous global optimisation combines a local search with Grover's algorithm,
and can locate the global optimum using effort proportional to the square root of the number of basins.
This article establishes that Jordan's quantum gradient estimation method can be incorporated into the quantum basin hopper,
providing an extra acceleration proportional to the domain dimension.
\end{abstract}

\noindent{\bf MSC2000 Subject Classification:}\ 90C30, 68Q99, 68Q25\\[0mm]

\def\fragment{\par\noindent\hrulefill\par}

\section{Introduction}

Recent articles~\cite{RGAS,GAS,DetH,Gocwin,NayakWu}
have studied the application of Grover's quantum algorithm~\cite{Boyer,Grover}
to the general unconstrained global optimisation problem.
Given a domain point and corresponding objective function value,
we can dynamically construct a quantum oracle classifying any other domain point as either better or not.
Using this oracle in Grover's algorithm produces an implementation of pure adaptive search~\cite{ZeS},
that is, an improving sequence of domain points, each uniformly distributed among all points better than the previous.
Details in~\cite{RGAS} show that the effort required by this approach
is on the order of the square root of the effort required by pure random search~\cite{Brooks}.

Of course, for many problems, more sophisticated methods than pure random search can be used on a Turing computer.
Typically in these methods the idea is to be ``guided'' toward optimal domain points by the objective function's ``terrain'',
that is, the shape of its graph.
For instance, the multistart method is a simple hybrid of pure random search and a local descent search,
in which independent exhaustive local searches are conducted from many randomly chosen initial points.
The quantum basin hopper~\cite{QBH}, a quantum analogue of the multistart method,
combines the quadratic acceleration due to Grover's algorithm
with the potentially huge acceleration over pure random search to be gained by using a local search.

In practice, objective functions in continuous domains are usually differentiable, at least piecewise.
Being guided by the terrain, in this case, will largely mean responding to the gradient,
an analytic expression for which may or may not be available.
Local optimisation techniques for such problems generally fall into two categories.
Gradient methods, such as gradient descent and conjugate gradient,
require explicit computation of the gradient at each sample point.
On the other hand, gradient-free methods, such as the Nelder-Meade simplex method
and simulated annealing, attempt to respond to the gradient in an indirect way,
without requiring its explicit computation.  (The methods named are all described in~\cite{Recipes}.)
Much of the reason for gradient-free methods is that,
when an analytic expression is available for the objective function but not for its derivative,
the number of function evaluations required near a point to estimate the gradient increases linearly with the domain dimension.
In many problems of practical interest, the domain dimension is large enough to make this unpractical.

The {\em quantum\/} complexity of estimating an objective function's gradient,
based solely on function evaluations, is constant, rather than linear, in dimension.
Specifically, \cite{Jordan} shows that the gradient can be estimated at a point
using just one (superposed) evaluation of the objective function.
In~\cite{GEA} it is established, under mild regularity conditions,
that the error in the gradient estimate can be reduced to any required level,
by appropriately selecting the numerical encoding and calculation precision
and other parameters of the gradient estimation algorithm.

Therefore it seems that gradient information is cheap in quantum computation,
and ought to be incorporated into local optimisation methods, and into the quantum basin hopper.
This paper discusses that possibility.
The difficulty is that, while~\cite{QBH} assumes the local search to act deterministically,
quantum gradient estimation produces in general a nontrivial superposition of computational basis states.
This is not simply a matter of numerical error;
the indeterminacy, together with the marking oracle used in the Grover search,
causes unwanted entanglement between the search register and the gradient estimation register.
We will argue that the error bound established in~\cite{GEA} implies that
the algorithm's performance is substantially unaffected by this entanglement.

This paper's structure is as follows.
Section~\ref{secQBH} presents the required details of the quantum basin hopper.
In Section~\ref{secErrorTypes},
we consider the error introduced by the indeterminacy of quantum gradient estimation,
and conceptually divide it into two different kinds of error, ``large'' and ``small''.
Section~\ref{secLarge} shows that the amplitude of the ``large'' errors can be made too small to disrupt the quantum basin hopper.
Section~\ref{secSmall} shows that the numerical scale of the ``small'' errors adequately restricts the effect of the entanglement.
Section~\ref{secConcl} concludes the paper and speculates on related future research.

\section{The quantum basin hopper}  \label{secQBH}

The quantum basin hopper algorithm is described in~\cite{QBH}.
The specifically quantum part of the algorithm, denoted GBS$(Y,r,L)$ below,
is nested within two loops which can be executed classically.

\begin{pseudocode}{Quantum basin hopper with parameters $(\lambda,M,L)$}
randomly choose $x\in S$
descend from $x$ to a local minimum $x_{cand}$
$y_{cand}\from f(x)$
{\bf repeat}
\	$x\from x_{cand}$
\	$y\from y_{cand}$
\	$m\from1$
\	{\bf repeat}
\	\	choose an integer $r$ uniformly at random from $\{0,\ldots,\lceil m-1\rceil\}$
\	\	set $x_{cand}$ and $y_{cand}$ by measuring the output of GBS$(Y,r,L)$
\	\	$m\from\lambda m$
\	{\bf until} $y_{cand}<Y$ {\bf or} $m>M$
{\bf until} $y_{cand}\geq Y$
\end{pseudocode}

The quantum part GBS$(Y,r,L)$ simply uses Grover's algorithm with $r$ rotations
to seek a domain point whose containing basin extends below $Y$.
(Basins are exlored using $L$ iterations of the local search.)
GBS$(Y,r,L)$ operates on a Hilbert space ${\cal D}^{\ox(L+1)}\ox{\cal R}$,
encoding the state of a system containing $L+1$ registers storing points in the domain of $f$,
and one register storing a value of $f$.  It proceeds as follows.
The state is initialised to $\ket s\ox\ket{0_{\cal D}}^{\ox L}\ox\ket{0_{\cal R}}$, where $\ket s$ is the equal-amplitude state in $\cal D$.
Then, the operator
\begin{equation}  \label{exprGBS}
U_f\of U_{d,L}\of\cdots\of U_{d,1}\of(U_s\of U_{d,1}^{-1}\of\cdots\of U_{d,L}^{-1}\of U_<\of U_f\of U_{d,L}\of\cdots\of U_{d,1})^r
\end{equation}
is applied.  Finally, the last two registers are measured,
and their values are returned as $x_{cand}$ and $y_{cand}$.
The factors process data as follows:
\begin{eqnarray*}
U_f\ket{x_0}\ox\cdots\ox\ket{x_L}\ox\ket{y} & = & \ket{x_0}\ox\cdots\ox\ket{x_L}\ox\ket{y+f(x_L)}, \\
U_s\ket{x_0}\ox\cdots\ox\ket{x_L}\ox\ket{y} & = & ((2\proj s-I)\ket{x_0})\ox\cdots\ox\ket{x_L}\ox\ket{y}, \\
U_<\ket{x_0}\ox\cdots\ox\ket{x_L}\ox\ket{y} & = &
 \left\{\begin{array}{ll} \ket{x_0}\ox\cdots\ox\ket{x_L}\ox\ket{y}, & y\geq Y, \\ -\ket{x_0}\ox\cdots\ox\ket{x_L}\ox\ket{y}, & y<Y, \end{array}\right. \\
\end{eqnarray*}
(where, for unitarity, the addition $y+f(x_L)$ is modulo the ordinate register's storage capacity).
The operator $U_{d,k}$ updates the $(k+1)$th register on the basis of the $k$th,
representing a single iteration of the local minimisation.
In~\cite{QBH}, it is assumed to act deterministically,
so that $U_{d,L}\of\cdots\of U_{d,1}$ maps the computational basis state
$$
\ket{x_0}\ox\ket{0}\ox\cdots\ox\ket{0}\ox\ket0
$$
to the computational basis state
$$
\ket{x_0}\ox\ket{x_1}\ox\cdots\ox\ket{x_L}\ox\ket0,
$$
where $x_1, \ldots, x_L$ is the local search's sample path when started at $x_0$.

\section{Errors due to quantum indeterminacy in gradient estimates}  \label{secErrorTypes}

In this article though, we assume instead that each operation $U_{d,k}$ involves a single quantum gradient estimation,
so that the operation~(\ref{exprGBS}) involves $(2r+1)L$ of them.
The effect of the indeterminacy inherent in the gradient estimation method on Grover's algorithm requires contemplation.
The terminus of the local minimisation sample path resulting from each domain point will be a quantum superposition,
peaked around a point near the bottom of the point's containing basin,
but having a nonzero amplitude at every domain point in general.
Accordingly, the marking oracle involved in Grover's algorithm,
which is here intended to identify points in basins extending below $Y$,
will always slightly entangle the register storing the sample path's origin with the circuitry estimating the gradient.

The aim of this article is to demostrate that Grover's algorithm, and therefore the quantum basin hopper,
can still function despite this entangling.
This will be done by considering separately the effects of ``large'' errors and ``small'' errors in the gradient estimates.
The error in a gradient estimate is simply the magnitude of the difference between the estimate and the true gradient,
and it is considered ``large'' if it exceeds $\delta$, and ``small'' otherwise,
where $\delta$ is a positive constant of our choosing.

\section{Large errors}  \label{secLarge}

If $\ket{g_n}$ is the $n$th gradient estimate,
let $\ket{g'_n}$ denote the normalised projection of $\ket{g_n}$
onto the span of estimates with error no larger than $\delta$.
Theorem~1 of~\cite{GEA} implies that, for any positive $\epsilon$ and $\delta$,
sufficiently careful quantum gradient estimation will ensure that $\|\ket{g'_n}-\ket{g_n}\|<\epsilon$.
Let $\ket\phi$ represent the computer's state just before measurement at the end of GBS$(Y,r,L)$,
and let $\ket{\phi'}$ represent the corresponding state if the gradient estimates $\ket{g_n}$
are replaced with $\ket{g'_n}$.  Clearly $\|\ket{\phi'}-\ket\phi\|<(2r+1)L\epsilon$,
which can be made as small as required.
Therefore 
we can safely ignore large errors.

\section{Small errors}  \label{secSmall}

It remains only to show that the quantum basin hopper can tolerate small errors,
that is, that if gradient estimate errors were uniformly bounded by a sufficiently small $\delta$,
the quantum calculation GBS$(Y,r,L)$ would locate a basin extending lower than $Y$ with a high probability
(for suitable rotation count $r$, and assuming such basins exist).

Let $f:D\to\Reals$ be the objective function before discretisation.
Suppose $D\subset\Reals^p$, and let $(D, {\cal F}, \mu)$ be a measure space with $\mu(D)$ finite and $f$ measurable.

Let $d_L(x_0, u_1, \ldots, u_L)$ denote the $L$th iterate of the local search,
starting from $x_0$, and assuming that the $k$th gradient estimate includes an error of $u_k$.
Let
\begin{eqnarray*}
W_+ & = & \{x\suchthat d_L(x,0,\ldots,0)>Y\}, \\
W_Y & = & \{x\suchthat d_L(x,0,\ldots,0)=Y\}, \mbox{ and} \\
W_- & = & \{x\suchthat d_L(x,0,\ldots,0)<Y\}.
\end{eqnarray*}
For almost all $Y$, $W_Y$ will have measure zero.
Ordinarily, for a wide class of local optimisation methods,
$d_L$ is continuous in all its arguments, almost everywhere.
Thus, letting $B_\delta$ represent the closed ball around $0\in\Reals^p$ with radius $\delta$, the set
$$
W_{Y,\delta} = \{x\suchthat d_L(x,B_\delta, \ldots, B_\delta) \mbox{ contains elements less than and greater than } Y\}
$$
decreases, as $\delta\to0$, to $W_Y$.
This means that we can make $W_{Y,\delta}$ as small as we like by choosing $\delta$ small enough,
that is, by computing the gradient estimates with sufficient precision.

Now suppose $S$ discretises $D$, so that $\cal D$ is spanned by $\{\ket x\suchthat x\in S\}$.
Partition $S$ into $S_\alpha$, $S_\beta$ and $S_\gamma$, where 
\begin{eqnarray*}
S_\beta & = & S\cap W_{Y,\delta}, \\
S_\alpha & = & (S\bs S_\beta)\cap W_+, \\
S_\gamma & = & (S\bs S_\beta)\cap W_-.
\end{eqnarray*}

Section~\ref{secQBH} indicates that the quantum search involves $L+1$ domain-point registers and one range value register;
in practice, there will be extra storage for ``scratch space'',
including the register for the gradient estimation sampling grid.
Taking all of this circuitry in account,
define the Hilbert space $\cal O$ such that $\cal D\ox O$ represents the state all registers involved in the quantum search,
with $\cal D$ representing the register storing $x_0$, and $\cal O$ representing all other registers.
Let $\ket{0_{\cal O}}$ represent the initialised state of $\cal O$.

Let
\begin{eqnarray*}
N_\alpha=|S_\alpha|, & & \ket\alpha=\sum\{\ket x\ox\ket{0_{\cal O}}\suchthat x\in S_\alpha\}/\sqrt{N_\alpha}, \\
N_\beta=|S_\beta|, & & \ket\beta=\sum\{\ket x\ox\ket{0_{\cal O}}\suchthat x\in S_\beta\}/\sqrt{N_\beta}, \\
N_\gamma=|S_\gamma|, & & \ket\gamma=\sum\{\ket x\ox\ket{0_{\cal O}}\suchthat x\in S_\gamma\}/\sqrt{N_\gamma}.
\end{eqnarray*}
Let
$$
G = U_s\of U_{d,1}^{-1}\of\cdots\of U_{d,L}^{-1}\of U_<\of U_f\of U_{d,L}\of\cdots\of U_{d,1},
$$
so that GBS$(Y,r,L)$ simply applies
$U_f\of U_{d,L}\of\cdots\of U_{d,1}\of G^r$
to $\ket s\ox\ket{0_{\cal O}}$.

Then
\begin{eqnarray*}
G\ket\alpha & = & \frac{N_\alpha - N_\beta - N_\gamma}N\ket\alpha +\frac{2\sqrt{N_\alpha N_\beta}}N\ket\beta + \frac{2\sqrt{N_\alpha N_\gamma}}N\ket\gamma
\mbox{ and} \\
G\ket\gamma & = & -\frac{2\sqrt{N_\alpha N_\gamma}}N\ket\alpha -\frac{2\sqrt{N_\beta N_\gamma}}N\ket\beta + \frac{N_\alpha+N_\beta-N_\gamma}N\ket\gamma;
\end{eqnarray*}
however, $G\ket\beta$ is not a tensor multiple of $\ket{0_{\cal O}}$, and represents an entangled state.
Let $\ket{\beta_k}$ denote $G^k\ket\beta$.

Define $\theta,\eta,\zeta\in[0,\pi/2]$ by
$$
\cos^2\theta = \frac{N_\beta^2}{4N_\alpha N_\gamma} \hskip48pt
\cos^2\eta = \frac{(N_\alpha-N_\gamma)^2}{(N_\alpha+N_\gamma)^2 - N_\beta^2} \hskip48pt
\cos^2\zeta = \frac{(N_\alpha+N_\beta/2)^2}{NN_\alpha}.
$$
Define
$$
\ket{\chi_+} = \frac1{\sqrt2}\left(e^{i\theta}\ket\alpha-\ket\gamma\right) \hskip64pt
\ket{\chi_-} = \frac1{\sqrt2}\left(e^{-i\theta}\ket\alpha-\ket\gamma\right).
$$
The Grover operation (not exactly a rotation in this context) acts on $\ket{\chi_\pm}$ as follows:
\begin{eqnarray*}
G\ket{\chi_+} & = & \frac1{\sqrt2}U_s\of U_{d,1}^{-1}\of\cdots\of U_{d,L}^{-1}\of U_<\of U_f\of U_{d,L}\of\cdots\of U_{d,1}\left(e^{i\theta}\ket\alpha-\ket\gamma\right) \\
& = & \frac1{\sqrt2}U_s\left(e^{i\theta}\ket\alpha+\ket\gamma\right) \\
& = & \frac{\ket\alpha}{\sqrt2}\left(\frac{2e^{i\theta}N_\alpha}N+\frac{2\sqrt{N_\alpha N_\gamma}}N-e^{i\theta}\right)
 + \frac{\ket\gamma}{\sqrt2}\left(\frac{2e^{i\theta}\sqrt{N_\alpha N_\gamma}}N+\frac{2N_\gamma}N-1\right) \\
 & & {} + \frac{\sqrt{2N_\beta}}N\left(e^{i\theta}\sqrt{N_\alpha}+\sqrt{N_\gamma}\right)\ket\beta \\
& = & \left(1 - \frac{2\sqrt{N_\gamma}}N\left(\sqrt{N_\gamma}+e^{i\theta}\sqrt{N_\alpha}\right)\right)\ket{\chi_+}
 + \frac{\sqrt{2N_\beta}}N\left(e^{i\theta}\sqrt{N_\alpha}+\sqrt{N_\gamma}\right)\ket\beta \\
& = & re^{-i\eta}\ket{\chi_+} + \frac{\sqrt{2N_\beta}}N\left(e^{i\theta}\sqrt{N_\alpha}+\sqrt{N_\gamma}\right)\ket\beta,
 \mbox{ where} \\
r & = & \sqrt{\frac{N_\alpha-N_\beta+N_\gamma}N}, \mbox{ and similarly} \\
G\ket{\chi_-} & = &
 re^{i\eta}\ket{\chi_-} + \frac{\sqrt{2N_\beta}}N\left(e^{-i\theta}\sqrt{N_\alpha}+\sqrt{N_\gamma}\right)\ket\beta.
\end{eqnarray*}
Therefore
\begin{eqnarray*}
G^k\ket{\chi_+} & = & r^ke^{-ik\eta}\ket{\chi_+}
 + \frac{\sqrt{2N_\beta}}N\left(e^{i\theta}\sqrt{N_\alpha}+\sqrt{N_\gamma}\right)
 \sum_{j=0}^{k-1}r^je^{-ij\eta}\ket{\beta_{k-1-j}} \mbox{ and} \\
G^k\ket{\chi_-} & = & r^ke^{ik\eta}\ket{\chi_-}
 + \frac{\sqrt{2N_\beta}}N\left(e^{-i\theta}\sqrt{N_\alpha}+\sqrt{N_\gamma}\right)
 \sum_{j=0}^{k-1}r^je^{ij\eta}\ket{\beta_{k-1-j}}.
\end{eqnarray*}
Noting that the initial state $\ket s\ox\ket{0_{\cal O}}$ equals
$$
-\frac i{\sqrt{2N}\sin\theta}(\sqrt{N_\alpha}+e^{-i\theta}\sqrt{N_\gamma})\ket{\chi_+}
 + \frac i{\sqrt{2N}\sin\theta}(\sqrt{N_\alpha}+e^{i\theta}\sqrt{N_\gamma})\ket{\chi_-}
 + \sqrt{\frac{N_\beta}N}\ket\beta,
$$
we have
\begin{eqnarray*}
G^k\ket s\ox\ket{0_{\cal O}} & = & \frac{ir^k}{\sqrt{2N}\sin\theta}
 \left(-e^{-ik\eta}(\sqrt{N_\alpha}+e^{-i\theta}\sqrt{N_\gamma})\ket{\chi_+}
 + e^{ik\eta}(\sqrt{N_\alpha}+e^{i\theta}\sqrt{N_\gamma})\ket{\chi_-}\right) + \sqrt{\frac{N_\beta}N}\ket{\beta_k} \\
 & & {} + \frac{i\sqrt{N_\beta}}{N\sqrt N\sin\theta}\sum_{j=0}^{k-1}r^j
 \left((\sqrt{N_\alpha}+e^{i\theta}\sqrt{N_\gamma})^2e^{i(j\eta-\theta)}
 - (\sqrt{N_\alpha}+e^{-i\theta}\sqrt{N_\gamma})^2e^{i(\theta-j\eta)}\right)\ket{\beta_{k-1-j}} \\
& = & \frac{r^k}{\sqrt{N}\sin\theta}
 \left((\sqrt{N_\alpha}\sin(k\eta)+\sqrt{N_\gamma}\sin(k\eta+\theta))\ket\gamma
 - (\sqrt{N_\alpha}\sin(k\eta-\theta)+\sqrt{N_\gamma}\sin(k\eta))\ket\alpha\right) \\
 & & {} + \sqrt{\frac{N_\beta}N}\ket{\beta_k} \\
 & & {} + \frac{i\sqrt{N_\beta}}{N\sqrt N\sin\theta}\sum_{j=0}^{k-1}r^j
 \left((\sqrt{N_\alpha}+e^{i\theta}\sqrt{N_\gamma})^2e^{i(j\eta-\theta)}
 - (\sqrt{N_\alpha}+e^{-i\theta}\sqrt{N_\gamma})^2e^{i(\theta-j\eta)}\right)\ket{\beta_{k-1-j}} \\
& = & \frac{r^k}{\sin\theta}\left(\sin(k\eta+\zeta)\ket\gamma - \sin(k\eta+\zeta-\theta)\ket\alpha\right) \\
 & & {} + \sqrt{\frac{N_\beta}N}\left(\ket{\beta_k}
 + \frac2{\sin\theta}\sum_{j=0}^{k-1}r^j\sin(\theta-2\zeta-j\eta)\ket{\beta_{k-1-j}}\right).
\end{eqnarray*}
In particular, noting that $\ket\gamma$ is orthogonal to $\ket\alpha$ and $\ket\beta$ but not necessarily to
$\ket{\beta_k}$ for $k>0$, we have
\begin{eqnarray}
\braopket\gamma{G^k}s & \geq & \frac{r^k\sin(k\eta+\zeta)}{\sin\theta}
 - \frac2{\sin\theta}\sqrt{\frac{N_\beta}N}\sum_{j=0}^{k-1}r^j|\sin(\theta-2\zeta-j\eta)| \label{ineqInterference} \\
& = & \frac{r^k\sin(k\eta+\zeta)}{\sin\theta}
 - \frac2{\sin\theta}\sqrt{\frac{N_\beta}N}\sum_{j=0}^{k-1}r^j\sin(\theta-2\zeta-j\eta) \mbox{ for small } k \nonumber \\
& = & \frac{r^k\sin(k\eta+\zeta)}{\sin\theta}
 - \frac i{\sin\theta}\sqrt{\frac{N_\beta}N}
 \left(e^{i(2\zeta-\theta)}\frac{1-r^ke^{ik\eta}}{1-re^{i\eta}}
 - e^{-i(2\zeta-\theta)}\frac{1-r^ke^{-ik\eta}}{1-re^{-i\eta}}\right). \nonumber
\end{eqnarray}
This bound is maximised when
$$
\frac{2\sqrt{N_\beta}}{\sqrt N}\Re\left(e^{i(2\zeta-\theta+k\eta)}\frac{\eta-i\ln r}{1-re^{i\eta}}\right)
 = \ln r\sin(k\eta+\zeta)+\eta\cos(k\eta+\zeta),
$$
that is, when
$$
k = \frac1\eta\atn\frac{2\sqrt{N_\beta/N}\rho\cos(2\zeta-\theta+\phi) - \ln r\sin\zeta - \eta\cos\zeta}
{2\sqrt{N_\beta/N}\rho\sin(2\zeta-\theta+\phi) + \ln r\cos\zeta - \eta\sin\zeta},
$$
where
$$
\rho e^{i\phi} = \frac{\eta - i\ln r}{1 - re^{i\eta}}.
$$
This reduces to the familiar situation when $N_\beta=0$; note that then we have
$$
\def\gap{\kern36pt}
\theta=\frac\pi2, \gap
\cos\eta = 1-2\frac{N_\gamma}N, \gap
\zeta = \eta/2, \gap
r = 1,
$$
and the optimal $k$ is $\pi/2\eta - 1/2$.

More tractibly, from~(\ref{ineqInterference}) we have
$$
\braopket\gamma{G^k}s \geq \frac1{\sin\theta}\left(r^k\sin(k\eta+\zeta)-2k\sqrt{N_\beta/N}\right).
$$
If $N_\beta\ll N_\gamma \ll N_\alpha$, then the usual estimate $k=(\pi/4)\sqrt{N/N_\gamma}$
gives
$$
\frac1{\sin\theta}\left(r^k\sin(k\eta+\zeta)-2k\sqrt{N_\beta/N}\right)
\approx
1-\frac\pi4\sqrt{\frac{N_\beta}{N_\gamma}}\left(1+\sqrt{\frac{N_\beta}N}\right);
$$
thus if the number of iterations is chosen optimally, the error rate is on the order of $\sqrt{N_\beta/N_\gamma}$.
This will usually be overshadowed by errors due to unknown target size,
which are already known to be manageable~\cite{Boyer}.

To summarise, ``small'' errors complicate the required implementations
of Grover's algorithm by introducing an entangling subset of the search space.
In the usual search algorithm, the search space is partitioned into marked and unmarked regions;
here we have instead a partition into marked, unmarked and entangling regions\footnote{The good, the bad and the ugly.}.
It is established here that if the entangling region is much smaller than the marked region,
then the search algorithm will still succeed with a high probability.

\section{Conclusion and further work}  \label{secConcl}

Taking, as a performance baseline, the pure random search algorithm~\cite{Brooks}
(simply, randomly sampling the domain repeatedly), the multistart method provides a large improvement,
for a wide class of optimisation problems.
Separately, implementing pure adaptive search~\cite{ZeS,FinitePAS} via Grover's algorithm~\cite{RGAS,GAS,DetH,Grover}
provides an acceleration on the order of the square root of the domain size;
this is impressive for a method not relying on any problem structure,
but quickly lags behind multistart for moderately structured problems.
In~\cite{QBH} it was shown how the quadratic acceleration due to Grover's algorithm,
and the acceleration (in proportion to the number of domain points in the deepest basin) due to multistart,
could be achieved simultaneously.

It is now known that the quantum complexity of gradient estimation is smaller than its classical complexity
by a factor which is linear in domain dimension~\cite{GEA,Jordan}.
The question naturally arises whether we can simultaneously claim the benefit of this linear acceleration,
alongside the accelerations due to Grover's algorithm and to multistart.
This paper provides the answer.
Section~\ref{secQBH} describes the quantum basin hopper algorithm.
Sections~\ref{secErrorTypes} to~\ref{secSmall} study the error introduced into the algorithm
by the indeterminacy inherent in quantum gradient estimation, using a ``divide-and-conquer'' approach.
Section~\ref{secErrorTypes} splits the error in gradient estimation into ``large'' and ``small'' errors.
Section~\ref{secLarge} shows that ``large'' errors are unproblematic due to their low amplitude.
Section~\ref{secSmall} shows that ``small'' errors only corrupt Grover's algorithm on a small subset of the search space,
with only slight overall effect.

To pursue this work further, a natural first step would seem to be to tighten the bounds established,
in order to predict, for a given optimisation problem,
what algorithm parameters and computational precision should be used.
A careful analysis of the performance of a quantum-gradient-estimate implementation of, say,
a conjugate gradient method on a quadratic form might yield insights.

\end{document}